\documentclass[12pt,letterpaper]{article} 
\usepackage{amssymb}
\usepackage{amsmath}
\usepackage{color}
\usepackage{fullpage}

 \newcommand{\bea}{\begin{eqnarray}}
\newcommand{\eea}{\end{eqnarray}}
\newcommand{\be}{\begin{equation}}
\newcommand{\ee}{\end{equation}}
\newcommand{\ba}{\begin{align}}
\newcommand{\ea}{\end{align}}

\newcommand{\RR}{\mathbb{R}} 

\newcommand{\HH}{\mathbb{H}}
\newcommand{\tr}{ ~{\rm Tr}}
\newcommand\rref[1]{(\ref{#1})}

\begin{document}

\title{\bf{A New Instability of the Topological black hole}}

\author{Alexandre Belin, Alexander Maloney}
\maketitle
\begin{center}
\ {\it  Physics Department, McGill University,\\ 3600 Rue University, Montr\'eal, QC H3A 2T8, Canada } \\ \smallskip
\vspace{1em} 
\texttt{alexandrebelin1986@gmail.com, maloney@physics.mcgill.ca}
\end{center}
\vspace{3em}

We investigate the stability of massless topological black holes in $AdS_d$ when minimally coupled to a scalar field of negative mass-squared.  In many cases such black holes are unstable even though the field is above the BF bound and the geometry is locally AdS.  The instability depends on the choice of boundary conditions for the scalars: scalars with non-standard (Neumann) boundary conditions tend to be more unstable, though scalars with standard (Dirichlet) boundary conditions can  be unstable as well.   This leads to an apparent mismatch between boundary and bulk results in the Vasiliev/Vector-like matter duality.

\newpage

\section{Introduction}

Black holes in anti-de Sitter space have been widely studied in the context of the AdS/CFT correspondence \cite{AdS/CFT,magoo}, where they are dual to finite temperature states. Their dynamical and thermodynamical stability properties therefore provide a novel window on the phase structure of the dual CFT. For example, in large $N$ gauge/gravity dualities the Hawking-Page phase transition in the bulk is dual to the confinement/deconfinement transition of the boundary theory \cite{deconfinement}.  Similarly, in holographic approaches to condensed matter physics the instability of a black hole to the condensation of scalar hair is dual to a superconducting phase transition \cite{superconductor1}\cite{superconductor2}.  

This paper focuses on the instability of topological black holes in AdS${}_d$, which were originally constructed in \cite{topbh} .  These black holes are quotients of global Anti-de Sitter space by a discrete group.  If the group is chosen appropriately, then the horizon $M^{d-2}$ will be a smooth space of constant negative curvature.  These black holes are the higher dimensional versions of the BTZ black hole in AdS${}_3$.  The horizon $M^{d-2}$ can be any hyperbolic manifold; for example, in four space-time dimensions the horizon can be any smooth Riemann surface of genus $g\ge 2$.  We will be interested in the case where the horizon topology is compact.
These solutions have been used to study behind-the-horizon physics in \cite{behindhorizon} and are important in the context of D-brane emission \cite{Barbon:2010us,Seiberg:1999xz,Buchel:2004rr,Kleban:2004bv,Maldacena:1998uz}.

These black holes are dual to CFT states on the surface $M^{d-2}$ at a particular value of the temperature related to the length scale of $M^{d-2}$.  It is possible to find other solutions where the temperature can be varied independently \cite{topbh}.  These are the negative curvature analogues of the Schwarzschild solution with finite mass \cite{topbh}.  We will focus on the "massless" case, which is locally AdS, where the geometry is relatively simple and it is possible to make precise analytic statements.  The stability of these black holes with respect to metric perturbations were studied in \cite{birmingham} (using the formalism of \cite{ishibashi}), where they were shown to be stable.  Similar studies were performed in \cite{Dias, GibbonsHartnoll, Wang:2001tk,Koutsoumbas:2008yq,Aros:2002te}, although the particular scalar instability that we will present here was not discussed. 

Our goal here is to generalize this analysis to consider other sorts of matter in the bulk.  We will focus on scalar fields of negative mass-squared.  In section 2 we will show that there is a range of values of $m^2$ where global AdS is stable, but the topological black hole is unstable.  This is despite the fact that the topological black hole is locally AdS; the essential observation is that certain modes which are typically discarded as non-normalizable in global AdS become normalizable once we perform the quotient to obtain the topological black hole. We will only discuss the linear instability and not its end-point. However, the authors of \cite{Dias} numerically constructed the backreacted solution for a particular mass in $AdS_5$ and we expect such hairy black holes to be the type of solutions found as the endpoints of these instabilities. 

The precise form of the instability will depend on which boundary conditions are chosen for the scalar field; these scalar fields are sufficiently light that two inequivalent quantization schemes can be chosen, corresponding to a choice of either standard (Dirichlet) or non-standard (Neumann) boundary conditions at asymptotic infinity.  For example, when $d>5$ the standard boundary conditions will be unstable if the scalar is sufficiently close to the BF bound.  Non-standard boundary conditions are even more unstable.  For example, In $d=4$, a scalar with mass $-2>m^2\ell^2>-5/4$ will lead to an instability if non-standard (Neumann) boundary conditions are imposed.  The marginal case $m^2 \ell^2 = -2$ is especially interesting, as it is a conformally coupled field in the bulk.  We will study these marginal cases separately, and conclude that they do not lead to instabilities because the associated modes have infinite energy.

These results have interesting implications for various boundary gauge theories on negatively curved spaces.
Our primary interest is the conjectured duality between bulk higher-spin Vasiliev gravity \cite{vasiliev} and gauge theories with vector matter, such as the $O(N)$ model, on the boundary \cite{Sezgin:2002rt,klebanovpolyakov,Giombi:2012ms} . This is the subject of section 3.  It is here that our focus on locally AdS solutions is useful; although the Vasiliev equations of motion are difficult to solve in general, AdS space is always a solution.  The equations of motion are local, so the massless  topological black holes is also a solution of Vasiliev theory.  Moreover, given the results above, the solutions appear to be stable in Vasiliev theory.  This is in contrast with the $O(N)$ model on a negatively curved compact surface.  The $O(N)$ scalars are conformally coupled so are effectively tachyonic, making the $O(N)$ model on a Riemann surface unstable.  This appears to contradict the proposed Vasiliev/$O(N)$ model duality, and at the very least indicates that some ingredients are missing in our current understanding of the duality.

{\bf Note Added}: We wish to emphasize that we are {\it not} saying that the $O(N)$ model on $\mathbb{R} \times \mathbb{H}_2$ is unstable; this space is conformal to Rindler space, so the theory will be stable.  Instead, the theory becomes unstable only when the spatial geometry is compactified to form a negatively curved Riemann surface. Similarly in the bulk, we are {\it not} claiming that AdS space in Rindler coordinates is unstable when the fields are above the BF bound; an instability will develop only when the spatial slices are compactified to form negatively curved compact spaces. The difference between the two scenarios relies on a detailed analysis of the scalar modes, and the distinction between which modes are viewed as normalizable on hyperbolic space, as opposed to normalizable on a compact Riemann surface. This analysis is the subject of this paper.


As an additional application of these results, in section 4 we consider the stability of various CFTs on compact negatively curved spaces.  For example, our results indicate that the CFT dual to $AdS_5 \times T^{1,1}$ will be unstable when the scalar has non-standard boundary conditions, corresponding to the addition of a particular double-trace operator in the CFT.

\section{Instability of the $AdS_d$ Topological Black Hole}

Topological black holes are solutions to Einstein's equation with a negative cosmological constant. In this section, we describe these black holes, solve the Klein-Gordon equation for a scalar field of arbitrary mass on this background, and look for modes that grow in time. We then discuss marginal cases as well as normalizability of the modes.

\subsection{The Topological Black Hole}

Topological black holes are solutions to Einstein's equations with a negative cosmological constant:
\be
R_{\mu\nu}=-\frac{d-1}{\ell^2}g_{\mu\nu} \ .
\ee
These black holes have horizons $M^{d-2}$, which are Einstein Manifolds with negative curvature. The metric is 
\be
ds^2=-f(r)dt^2+f^{-1}(r)dr^2+r^2H_{ij}dx^idx^j \label{generalmetric} \ ,
\ee
with
\be
f(r)=\left(-1-c\frac{\mu}{r^{d-3}}+\frac{r^2}{\ell^2}\right)
\ee
where c a constant chosen so that $\mu$ has dimensions of inverse length.
In fact, these black holes can have zero or negative mass. We will be interested here in the massless topological black hole that is locally $AdS_d$. Setting $\mu=0$, rescaling $t$ by a factor of $\ell$ and making the coordinate change $r=\ell\cosh\rho$, the metric becomes
\be
{ds^2\over \ell^2} =  - \sinh^2 \rho dt^2 + d\rho^2+\cosh^2 \rho dH_{d-2}^2 \label{metric} \ .
\ee
If $dH^2_{d-2}$ is the constant (unit) negative curvature metric on the $d-2$ dimensional hyperboloid $\HH_{d-2}$, this is just $AdS_d$ written in hyperbolic slicing. The asymptotic boundary is $\HH_{d-2}\times \RR_t$ in this coordinate system\footnote{ Note that $\HH_{d-2}$ itself has a boundary sphere $S^{d-3}$, which reflects the fact that this coordinate system does not cover all of $AdS_d$.  On a constant time slice one should really view the global boundary $S^{d-2}$ as two copies of $\HH_{d-2}$ which are glued together along their boundaries.}.

We will  be interested in geometries where $dH^2$ is replaced by $d\Sigma^2$, the constant negative curvature metric on a surface $\Sigma = \HH_{d-2}/\Gamma$.  Here $\Gamma$ is a freely acting subgroup of the isometry group $SO(d-2,1)$ of $\HH_{d-2}$, so that $\Sigma$ is smooth.  The surface $\Sigma$ can be taken to be either compact or non-compact and may have any genus $g$ for $g\geq2$; we will focus on the compact case, and consider arbitrary genus. 

\subsection{Scalar Field}

Let us now consider a minimally coupled scalar field $\phi$  which obeys the free wave equation
\be
(\nabla^2 - m^2) \phi = 0
\ee   
We wish to understand the stability of the quotient of $AdS_d$ described above.  We will expand $\phi$ in modes on $\Sigma$ as
\be
\phi = \tilde{R}_{\omega \lambda}(\rho) e^{\omega t} Y(\sigma)
\ee
where $Y(\sigma)$ is an eigenfunction of the scalar Laplacian on $\Sigma$: $\nabla_{\Sigma}^2 Y = -\lambda Y$.  
We will seek unstable solutions which grow in time, so that $\omega$ is real and positive. Fluctuating solutions would have $\omega$ imaginary while quasinormal modes would have $\omega$ complex. For certain values of the mass we will find solutions with $\omega=0$, indicating the presence of an additional logarithmic mode; in these cases the expansion in terms of exponentials $e^{\omega t}$ fails and a more detailed analysis is required, which will be the subject of the following subsection.

We will follow the analysis of \cite{birmingham}. Letting $\tilde{R}=(\cosh\rho)^{-\frac{d-2}{2}}R$ the radial equation for $R$  reduces to a hypergeometric equation with solutions
\be\label{Rhor}
R = C z^{\alpha}(1-z)^\beta F(a,b,c,z) + D z^{-\alpha} (1-z)^\beta F(a-c+1, b-c+1, c-2, z)
\ee
where $z=\tanh^2 \rho$, 
\be
\alpha = {\omega/2},~~~~~\beta = {1\over 4} - {1\over 4} \sqrt{(d-1)^2 + 4 m^2 \ell^2 }
\ee
and 
\bea
a &=& {1\over 2}+ {\omega/2}- {1\over 4} \sqrt{(d-1)^2 + 4 m^2 \ell^2 } + {1\over 4} \sqrt{(d-3)^2 - 4\lambda }\\
b &=& {1\over 2}+ {\omega/2}- {1\over 4} \sqrt{(d-1)^2 + 4 m^2 \ell^2 } - {1\over 4} \sqrt{(d-3)^2 - 4\lambda }\\
c&=& 1 + \omega
\eea

We are interested in solutions which are smooth in the interior, and in particular are smooth at the horizon ($\rho=0$).  At this point the hypergeometric functions in \rref{Rhor} become constant. Since we are assuming that $\omega$ is real and positive, this implies that we must set $D=0$ in \rref{Rhor}.  If $\omega$ were purely imaginary, this branch of solutions would describe an infalling wave at the horizon.

We now investigate the behavior near the asymptotic boundary. To do so, we use properties of the hypergeometric functions to transform \rref{Rhor} and obtain
\be
R \sim A e^{\left(-{1\over 2} + {1\over 2} \sqrt{(d-1)^2 + 4 m^2 \ell^2} \right) \rho}
+B e^{\left(-{1\over 2} - {1\over 2} \sqrt{(d-1)^2 + 4 m^2 \ell^2} \right) \rho} \label{asymptotics}
\ee
at $\rho\to\infty$.  Here 
\be
A = {\Gamma(c) \Gamma(c-a-b)\over \Gamma(c-a)\Gamma(c-b)},~~~~~
B = {\Gamma(c) \Gamma(a+b-c)\over \Gamma(a)\Gamma(b)},~~~~~
\ee
The second of these falls off more quickly at the asymptotic boundary than the first, giving us the two possible boundary conditions for a light scalar in $AdS_d$. Let us first identify perturbations with standard (Dirichlet) boundary condition $A=0$.  
Setting $A=0$ requires that either $c-a$ or $c-b$ is a non-positive integer.  That is, 
\be\label{Acond}
{1\over 2} + \omega/2 + {1\over 4} \sqrt{(d-1)^2 + 4 m^2 \ell^2}\pm {1\over 4} \sqrt{(d-3)^2 - 4 \lambda} =-n
\ee 
for some non-negative integer $n$.  Likewise perturbations obeying the non-standard (Neumann) boundary condition have $B=0$, which requires $a$ or $b$ to be a positive integer.  This requires
\be\label{Bcond}
{1\over 2} + \omega /2 - {1\over 4} \sqrt{(d-1)^2 + 4 m^2 \ell^2 } \pm {1\over 4} \sqrt{(d-3)^2 - 4\lambda }=-n
\ee
for a non-negative integer $n$.  We will consider fields whose masses obey the BF bound \cite{BF1,BF2}, $m^2 \ell^2> -(d-1)^2/4$, so that $\sqrt{(d-1)^2 + 4 m^2 \ell^2}$ is real.

Let us now investigate the stability of the $\phi=0$ vacuum.  We need to understand the spectrum of the Laplacian $\nabla^2_\Sigma Y = - \lambda Y$.  When $\Sigma$ is just the hyperboloid $\HH_{d-2}$, the normalizable modes form a continuum of states with $\lambda > (d-3)^2 / 4$.  Thus $\sqrt{(d-3)^2 -4 \lambda}$ is imaginary and the equations \rref{Acond} and \rref{Bcond} both require $\omega$ to be imaginary.  So the perturbations fluctuate periodically and the $\phi=0$ vacuum is stable.  

However, when $\Sigma=\HH_{d-2}/\Gamma$ is a non-trivial quotient the spectrum of the Laplacian may change.  In particular, normalizable modes with $\lambda < (d-3)^2/4$ may exist.  We consider the extreme case where $\Sigma$ is compact, in which case the constant mode with $\lambda=0$ is normalizable.  \rref{Acond} and \rref{Bcond} can now be solved with $\omega$ real and positive, which leads to instabilities.

Let us consider the constant mode on $\Sigma$ with standard boundary conditions $A=0$.  The most unstable mode will have $n=0$, and has frequency \footnote{We consider $d\geq4$ as $\mathbb{H}_1$ is just the line.}
\be
\omega = {1\over 2}\left(d-5-\sqrt{(d-1)^2 + 4 m^2 \ell^2}\right)
\ee
Notice we chose the - sign in \rref{Acond} as the + sign always leads to $\omega<0$. For $d=4,5$, we find no instabilities but when $d>5$ the standard boundary conditions will be {\it unstable} if 
\be
-(d-1)^2/4< m^2 \ell^2 <  -2(d-3) 
\ee
Note that masses in this range are {\it above} the BF bound thus even though global $AdS_d$ is stable, with these boundary conditions, the massless topological black hole is not.

Turning our attention to non-standard boundary conditions, we demand $B=0$.  These boundary conditions still define  normalizable modes (with respect to the Klein-Gordon norm) provided we are in the window
\be
-(d-1)^2/4 < m^2 \ell^2 < - (d-3) (d+1)/4 \label{normalizibility}
\ee
the lower bound being again the usual BF bound.
The most unstable mode with $n=0$ will have frequency
\be
\omega = {1\over 2}\left(d-5+\sqrt{(d-1)^2 + 4 m^2 \ell^2}\right)
\ee
When $d\geq5$, the topological black hole with non-standard boundary conditions will always be unstable provided that \rref{normalizibility} is satisfied. When $d=4$ the non-standard boundary condition is unstable when
\be
-2<m^2 \ell^2< - 5/4 \ .
\ee
Again, this means that for a large range of masses the non-standard quantization  of a scalar field in the massless topological black hole background is unstable, even though global $AdS_d$ is stable.


\subsection{The Marginal Case and Finite Energy}

As we have seen, the bound $m^2\ell^2=-2(d-3)$ appears both for the standard boundary condition ($d\geq5$) and for the non-standard boundary condition ($d=4,5$). We solve the wave equation in appendix \ref{marginalmass} and find modes that grow linearly in time. However, an additional issue needs to be taken into account. So far we have demanded that the modes are normalizable with respect to the Klein-Gordon norm but we must also demand finiteness of the stress-tensor associated to the modes.  It turns out all modes considered in the previous subsection have a finite energy.  However, as we will now show, the linearly growing modes appearing in the marginal case have infinite energy.

So far, we have only required  the radial functions of the field $\tilde{R}(\rho)$ to be regular at the horizon.  However, requiring the energy of these perturbations to be finite imposes a stronger condition. As explained in \cite{GibbonsHartnoll}, the differential equation we are solving for the modes can be reduced to a Schrodinger-like equation by going to Kruskal coordinates, which are well behaved at the horizon. The differential equation is
\begin{equation}
-\frac{d}{dr}\left(fr^{d-2}\frac{d\tilde R}{dr}\right)+\left(\frac{-\lambda}{r^2}-\frac{2f'}{r}-\frac{(2d-4)f}{r^2}-2(d-1)\right)r^d\tilde{R}=-\omega^2\frac{r^{d-2}}{f}\tilde{R}
\end{equation}
with $r=\ell \cosh\rho$ and $f(r)=-1+r^2/\ell^2$. We can now set
\begin{equation}
R=\tilde{R}r^{(d-2)/2} \ \ \ \ \ \ \ dr_*=\frac{dr}{f}
\end{equation}
The equation now becomes
\begin{equation}
-\frac{dR}{dr^2_*}+V(r(r_*))R=-\omega^2R\equiv ER \label{schrodinger}
\end{equation}
with
\begin{equation}
V(r)=\frac{-\lambda f}{r^2}+\frac{(d-6)f'f}{2r}+\frac{(d^2-14d+32)f^2}{4r^2}-2(d-1)f
\end{equation}
and $E$ the Schrodinger energy. We see that \rref{schrodinger} is a Schrodinger equation, which may have bound states with $E<0$. As we have seen in the previous subsection, the threshold value corresponding to $\omega=0$ is also interesting as it corresponds to linearly growing modes. We now turn our attention to the finiteness of the stress tensor associated to these perturbations. The energy of a given perturbation is
\be
E\propto \int \sqrt{g^{d+1}}d^{d+1}x T_{\mu\nu}n^\mu\xi^\nu
\ee
where $T_{\mu\nu}$ is the stress tensor of the perturbation, $g^{d+1}$ is the metric on the hypersurface of a fixed time slice, $n^\mu$ its normal vector and $\xi^\mu$ the timelike killing vector with $\xi^0=1$. As explained in \cite{GibbonsHartnoll}, finiteness of the stress tensor implies normalizability of the wave function. In other words, we must require that 
\begin{equation}
\int R^2dr_*=\int R^2\frac{dr}{f}=1
\end{equation}
Going back now to our $\rho$ coordinate for which we have solved the differential equation we see that this corresponds to
\begin{equation}
\int \frac{R^2(\rho)d\rho}{\sinh\rho}
\end{equation}
We now recall from \rref{Rhor} that $R\sim \tanh\rho^{\omega}\sim\sinh^{\omega}\rho$ near $\rho=0$.
As there clearly is no problem at infinity, the only potential problem for convergence of the integral is at the horizon. We see that as long as $\omega>0$, the integral will converge and the energy of the perturbation is finite. This requires that $R\rightarrow0$ at the horizon (as $f$ has a simple pole) which is the case for all our non-marginal modes. However, the marginal case we have considered in the previous subsection describes modes that are constant at the horizon. This means the perturbations associated with these modes have infinite energy and should not be considered.   As we will see in the next section, this presents us with a puzzle in the context of the Klebanov-Polyakov duality.

\section{Instability of the $O(N)$ Model on a Riemann Surface}

\subsection{Free Theory}

Probably the most interesting application of our results concern the Giombi-Klebanov-Polyakov-Yin duality \cite{Sezgin:2002rt,klebanovpolyakov,Giombi:2012ms} relating Vasiliev's higher spin theory with non-standard (standard) boundary conditions to the free (critical) O(N) model. As we will see shortly, our results lead to an apparent mismatch between the bulk and boundary dynamics. 

Let us begin by considering the free 2+1 dimensional $O(N)$ model on a compact Riemann surface $\Sigma$ of negative curvature $R<0$. The action is
\be
S =- {1\over 2} \int dt d\Sigma \tr \left((\nabla \Phi)^2 +{1\over 8} R \Phi^2 \right)  
\ee
This theory is unstable, since the conformal coupling acts as a negative mass squared.  The spin-0 current $J=\tr \Phi^2$ will condense as the components of $\Phi$ run off to infinity.  Note that this is an exponential instability; the vev of the field $\Phi$ increases exponentially in time. We wish to emphasize the importance of having a compact Riemann surface: if no quotient is performed the space is simply $\mathbb{R}\times \mathbb{H}_2$ and the theory cannot be unstable. Indeed, it is conformally related to Rindler space and a free scalar on Rindler space is perfectly stable. The crucial difference with a compact Riemann surface is that the constant mode of the field $\Phi$ becomes a valid normalizable mode and it is precisely this mode that will be responsible for the instability of the theory.

The GKPY duality really involves the singlet sector of the $O(N)$ model and the projection is done by coupling the scalar field to a Chern-Simons field. We note that on a manifold with non-trivial topology the Chern-Simons field could have non-trivial holonomy around a non-contractible cycle.  The choice of the holonomy changes the scalar kinetic term, and can in principle alter the zero-mode dynamics of the scalar \cite{lightstates}\cite{LargeNCSMatter}\cite{dSTopology}.  Thus the scalar dynamics on its own, for fixed holonomy, could be stable.  However, the holonomies themselves are dynamical, continuous degrees of freedom.  Thus the energetics of the theory will push the holonomies towards zero, at which point the scalar field is unstable. 

We can now investigate this instability from the bulk.  
The topological AdS black hole of the previous section is -- being locally AdS -- necessarily a solution of the Vasiliev equations of motion.  
This geometry should be the dual to the $O(N)$ model at finite temperature.
The operator $J$ has dimension $\Delta=1$ and corresponds to a bulk field $\phi$ of mass $m^2 \ell^2 = -2$ with non-standard boundary conditions.  However, according to the analysis of the previous section, we find no instability inside the bulk. The condition of finite energy ruled out the polynomial time dependence and thus we must conclude we find no instability at the linearized level.  

There are two possible resolutions to this apparent mismatch between the bulk and boundary dynamics.
One possibility is that there are new non-local dynamics in the bulk that arise when the geometry has nontrivial topology that could cause an instability.  Given that Chern-Simons theory on its own is dual to topological string theory \cite{Gopakumar:1998ki}, it is natural to expect that the duals of Chern-Simons-Matter theories could contain topological degrees of freedom.  The bulk theory may be sensitive to these degrees of freedom only once it is put on a manifold of non-trivial topology, in the same way that the holonomy degrees of freedom of a Chern-Simons theory become relevant on a manifold with non-contractible cycles.  
Our results would then indicate that these non-local degrees of freedom become active and unstable on the topological black hole.

It is also possible that other higher spin fields are unstable, even though the scalar is stable.  However, given that the boundary instability is visible in the dynamics of $J$, this would not resolve this puzzle; presumably the scalar operator would be unstable through it's couplings with higher spin currents, but this would be visible only at the non-linear level.


\subsection{Critical $O(N)$ and Free Fermion Models}

We can also consider the critical $O(N)$ model, which is dual to the same Vasiliev bulk theory, except that the scalar field is now given standard boundary conditions.  From the above analysis we see that the scalar field is stable.

Note that this may not be a contradiction, as the instability of the critical $O(N)$ model does not cause $J=\tr \Phi^2$ to run off to infinity.  Instead $J$ will condense at an order one value.  In this case we would expect the instability to be invisible in the bulk.  It would be good to work this out explicitly.  In particular, our conjecture is that the topological black hole will acquire scalar hair, but that the scalar hair comes with factors of N which render it invisible in the classical bulk analysis.

Likewise, the free fermion model is also stable, because the Dirac fermion is, by itself, conformally coupled without the need for additional terms that couple to curvature.  This is consistent with the fact that, unlike the scalar case, a fermion mass term would violate parity invariance in three dimensions.  We note that by level-rank duality, the critical fermion theory (i.e. the Gross-Neveu model) at $\lambda \sim \mathcal{O} (1)$ is dual to the free boson at small $\lambda$ \cite{Maldacena:2012sf,Aharony:2012nh,GurAri:2012is,Aharony:2012ns,Jain:2013py}.  Thus it seems likely that there is a critical value of $\lambda$ at which the instability of the free boson theory disappears.

\section{The instability of theories on hyperbolic spaces}
Having discovered these instabilities, we wish to look for further examples among the known AdS/CFT dualities. 
We start by reviewing the spectrum of Kaluza-Klein fluctuations in $AdS_5\times S^5$ -- which is stable -- and then discuss an unstable example, $AdS_5\times T^{1,1}$, where we can identify the unstable operators in the dual field theory.

\subsection{$AdS_5\times S^5$}
From the analysis done in section 2, we can set $d=5$ and see that standard boundary conditions will not lead to instabilities whereas non-standard boundary conditions will. The range of masses for unstable modes is
\begin{equation}\label{window}
-4<m^2\ell^2<-3
\end{equation}
We are interested in the spectrum of Supergravity on $AdS_5\times S^5$, given in \cite{ads5_s5}. We will not give the expansions of all fields in harmonics on $S^5$ but the fields of interest are the metric perturbations and the 4-form perturbations. The fields have indices living both on the $AdS$ space and the 5-sphere but we are interested in the components where all indices live on the 5-sphere ($AdS$ scalars):
\bea
h^a_a&=&\sum_l\pi^l(x)Y^l(y) \notag \\
a_{abcd}&=&\sum_l b^l(x)\epsilon_{abcd}^{\phantom{abcd}e}D_eY^l(y)
\eea
where we have transformed the symmetric traceless 4-tensor spherical harmonic into the scalar spherical one. One then solves the linearized equations and one finds a mixed system in $b^l$ and $\pi^l$
\be
\Box_x\begin{pmatrix} \pi \\ b \end{pmatrix}+\begin{pmatrix} \Box_y-32\ell^2 & 80\ell^2\Box_y \\ -\frac{4}{5}\ell & \Box_y \end{pmatrix}\begin{pmatrix} \pi \\ b \end{pmatrix}=0 \label{piandb}
\ee
where $\Box$ denotes the scalar laplacian and the mass is taken to be the eigenvalue of $\Box_x$. One then finds the two following branches
\bea
m^2\ell^2&=&l(l-4) \ \ \ \ \ \ \ \ \ \ \ (l\geq2) \notag \\
m^2\ell^2&=&(l+4)(l+8)  \ \ \ \ \  (l\geq0)
\eea
The branch we are interested in is the first one, as the first mode will have $m^2\ell^2=-4=m_{BF}$ which is also the marginal case studied in section 2.
 The first exited mode will have $m^2\ell^2=-3$ which is now too big to cause instabilities. We can also look at the mass branches in \cite{ads5_s5} and see that a component of the 2-form has a bulk scalar with $m^2\ell^2=-3$ but no field lies between $-3$ and $-4$.

By the above analysis, we see that the topological black hole solution of $AdS_5\times S^5$ is stable against scalar perturbation. We cannot conclude on global stability as we have done this analysis for scalars only but we can conclude that, to our knowledge, there is no instability. The other classical examples of $AdS^4\times CP^3$, $AdS^4\times S^7$ or $AdS_7\times S^4$ do not lead to instabilities either \cite{cp3, mq}.
%
\subsection{$AdS^5\times T^{1,1}$}
To find examples with scalars in the window (\ref{window}) we must consider theories with less supersymmetry, such as those discussed by Klebanov and Witten \cite{klebanov} describing D-branes at a Calabi-Yau singularity. We will consider the simplest case, where the corresponding bulk geometry is $AdS_5\times T^{1,1}$ with $T^{1,1}=\frac{SU(2)\times SU(2)}{U(1)}$ (in fact there are several ways to embed  $U(1)$ into $SU(2)\times SU(2)$ giving rise to the more general $Y^{pq}$). The mass spectrum was calculated in \cite{t11} --  we will extract the relevant information and summarize it below. 

The harmonic expansion on $T^{1,1}$ gives the following spectrum of the scalar Laplacian \cite{t11}\cite{gubser} 
\be
\Box Y^{(k,l,r)}=H_0(k,l,r)Y_0^{k,l,r}
\ee
with
\be
H_0(k,l,r)=6\ell^2\left(k(k+1)+l(l+1)-\frac{r^2}{8}\right)
\ee
where $k$ and $l$ correspond to the two $SU(2)$ quantum numbers and $r$ is the $R$-symmetry quantum number (it corresponds to the $U_R(1)$ charge whose generator is orthogonal to that of the quotient group $U_H(1)$, additional details are given in \cite{t11}). The two $SU(2)$ quantum numbers $k$ and $l$ must either both be integers or half-integers. The scalar spectrum is richer than that on $S^5$, and the scalars distribute themselves in the 9 different multiplets. By analyzing the mass spectrum, one quickly finds scalars in the interested mass range. One such example is again a linear combination of the $b$ and $\pi$ fields of the previous subsection, that follow a mixed equation of the type of \rref{piandb}.  This field lies in one of the vector multiplets \cite{t11}. For $k=l=1/2$ and $r=1$, the field has mass\footnote{There are of course other fields that have masses lying in the range of instability, we focus here on one simple example.}.
\be
\label{mode}
m_{b}^2\ell^2= -\frac{15}{4} \,.
\ee
We thus conclude that the topological black hole will be unstable for type IIB string theory compactified on $T^{1,1}$, provided that this scalar is given Neumann boundary conditions. 

From the CFT point of view this scalar is dual to the operator 
\be
\mathcal{O}=\tr A B
\ee
 where $A$ and $B$ are the bifundemental chiral superfields of the quiver gauge theory. They are respectively in the $(N,\bar{N})$ and $(\bar{N},N)$ representations of the $SU(N)\times SU(N) $ gauge group \cite{Baumann:2010sx}. 
Our bulk instability therefore leads to us to conjecture that, at strong coupling, the quiver gauge theory deformed by the double trace deformation $\mathcal{O}^2$ will have an instability on $R\times \mathbb{H}_3/\Gamma$ at temperature $\beta=2\pi$.  This instability will cause the bulk scalar to condense, so that the dual operator will acquire a non-zero expectation value.

\section{Acknowledgements}
We are grateful to S. Banerjee, G.~Horowitz, S.~Matsuura, S.~Ross, H.~Reall, E. Silverstein, V. Didenko, E. Skvortsov, M. Vasiliev, R. Eager and especially S. Shenker for useful discussions.  
AM is supported by the National Science and Engineering Research Council of Canada. AB is supported by the Swiss National Science Foundation.

\appendix
\section{Solving the wave equation for the marginal mass \label{marginalmass} }
In this appendix, we solve the wave equation for the marginal value of the mass $m^2\ell^2=-2(d-3)$\footnote{For now, we will not consider the case $d=5$ which involves an additional subtlety we leave for the following subsection.}. We look for solutions of the form:
\be
\phi = \tilde{R}_{\omega \lambda}(\rho) f(t) Y(\sigma)
\ee
The Klein-Gordon equation is
\be
\left(-\frac{\partial_t^2}{\sinh^{2}\rho}+(\coth\rho+(d-2)\tanh\rho)\partial_\rho+\partial_\rho^2-\frac{\lambda}{\cosh^{2}\rho}-m^2\ell^2\right)\tilde{R}f(t)=0
\ee
As the marginal mass led to $\omega=0$, the exponential time dependence is no longer an option. This means $f$ is either constant or linear in time. Thus the unstable mode will grow linearly in time. The solutions of the radial equation are now the following
\begin{equation}
R = C(1-z)^\beta F(a,b,c,z) + D (1-z)^\beta \left(F(a, b, 1, z)\ln z + \sum_{k=1}^{\infty}z^k\frac{(\alpha)_k(\beta)_k}{(k!)^2} \Psi(k)\right)
\end{equation}
where
\begin{equation}
\Psi(k)=\psi(\alpha+k)-\psi(\alpha)+\psi(\beta+k)-\psi(\beta)-2\psi(k+1)+2\psi(1) \nonumber
\end{equation}
\be
\psi(x) = \partial_x\ln\Gamma(x),~~~~~\beta = {1\over 4} - {1\over 4} |d-5|
\ee
and 
\bea
a &=& {1\over 2}- {1\over 4} |d-5| + {1\over 4} \sqrt{(d-3)^2 - 4\lambda }\\
b &=& {1\over 2}- {1\over 4}|d-5| - {1\over 4} \sqrt{(d-3)^2 - 4\lambda }
\eea
Again, we want a smooth function at the origin, so we need to impose $D=0$ because of the log in the second solution. To check the behavior at infinity, we do the same transformation as in \rref{asymptotics} and we get:
\be
R \sim A e^{\left(-{1\over 2} + {1\over 2} |d-5| \right) \rho}
+B e^{\left(-{1\over 2} - {1\over 2} |d-5|\right) \rho}  \label{asymptotics_marginal}
\ee
This time
\be
A = {\Gamma(1-a-b)\over \Gamma(1-a)\Gamma(1-b)},~~~~~
B = {\Gamma(a+b-1)\over \Gamma(a)\Gamma(b)},~~~~~
\ee
Let us start with the standard boundary conditions. As before, there is no constraint due to normalizibility as these modes are all normalizable and selecting the standard boundary conditions imposes $A=0$. The condition is then:
\be
{1\over 2}+ {1\over 4} |d-5| - {1\over 4} \sqrt{(d-3)^2 - 4\lambda }=-n
\ee
for a non-negative integer $n$. For $d<5$, this will never be satisfied. For $d>5$, this will always be satisfied for the constant monde on $\Sigma$ ($\lambda=0$) and there is a mode growing linearly in time.

Turning our attention to the non-standard boundary condition, we are interested in $d=4$. We must set $B=0$ in \rref{asymptotics_marginal} which gives us the condition:
\be
{1\over 2}- {1\over 4}  \pm {1\over 4} \sqrt{1 - 4\lambda }=-n
\ee
The + sign in the equation above (corresponding to $a=0$) is not interesting and we consider only the - sign.
We find that the condition is satisfied in $d=4$ for the constant mode so the quantization of the scalar field with non-standard boundary conditions is unstable.

Let us summarize the results for the marginal case. Imposing standard boundary conditions, we find linearly growing modes for $d>5$. Imposing non-standard boundary conditions in $d=4$, we also find linearly growing modes.

\subsection{the BF-scalar in $d=5$}
We would like to give a little more attention to the marginal case in five dimensions, which gives a mass
\be
m^2\ell^2=-4
\ee
corresponding both to our marginal case \textit{and} to a BF scalar. As we shall see in section 4, such a field appears in the supergravity spectrum on $AdS_5\times S^5$, which makes it of particular interest. Things get a little more complicated for this marginal example because \rref{asymptotics_marginal} is no longer true as what we previously called standard and non-standard boundary conditions would have the same exponential fall out and one needs to go back to the differential equation and solve it directly with the appropriate parameters. We find
\be
a=1 \ \ \ \ \ \ b=0 \ \ \ \ \ \ c=1 \ \ \ \ \ \ \alpha=0 \ \ \ \ \ \ \beta=\frac{1}{4}
\ee
The solution of the differential equation is
\be
R(z)=A+B(\log z - \log (1-z))
\ee
One can see right away that that the second solution will not be normalizable as it blows up at the origin $z=0$.
The asymptotics of the total radial function $\tilde{R}(\rho)$ is
\be
\tilde{R}(\rho)\sim e^{-2\rho}(A+b\rho^2)
\ee
The interpretation of standard or non-standard boundary conditions is more delicate here as they have the same fall off, up to some logarithmic function. We will call the one falling off quicker the standard boundary condition. In this case, the solution is particularly simple and there is no need to "transform" hypergeometric functions to find the asymptotic behavior. We see that non-standard boundary condition are ruled out by regularity at the horizon and only standard boundary conditions would have modes growing linearly in time.


\begin{thebibliography}{2}
\bibitem{AdS/CFT} J. Maldacena, \textit{The large N limit of superconformal field theories and supergravity} Adv. Theor. Math. Phys. \textbf{2} (1998) 231-252; hep-th/9711200
\bibitem{magoo} O. Aharony, S. Gubser, J. Maldacena, H. Ooguri and Y. Oz, \textit{Large N field theories, String Theory and Gravity}, hep-th/9905111
\bibitem{deconfinement} E. WItten, \textit{Anti-de Sitter Space, Thermal Phase Transition, And Confinement In Gauge Theories}; hep-th/9803131
\bibitem{superconductor1} S. A. Hartnoll, C. P. Herzog, G. T. Horowitz, \textit{Building an AdS/CFT supercon-
ductor}; hep-th/0803.3295
\bibitem{superconductor2} S. A. Hartnoll, C. P. Herzog, G. T. Horowitz, \textit{Holographic Superconductors}; hep-th/0810.1563
\bibitem{topbh}
  R.~Emparan,
  JHEP {\bf 9906}, 036 (1999)
  [hep-th/9906040].
\\
  R.~Emparan,
  Phys.\ Lett.\ B {\bf 432}, 74 (1998)
  [hep-th/9804031].
\\
 D. Birmingham, \textit{Topological Black Holes in Anti-de Sitter Space}, Class. Quant. Grav. \textbf{16} (1999) 1197; hep-th/9808032

\bibitem{behindhorizon} G. Horowitz, A. Lawrence and E. Silverstein, \textit{Insightful D-branes}; hep-th/0904.3922

\bibitem{Barbon:2010us} 
  J.~L.~F.~Barbon and J.~Martinez-Magan,
  JHEP {\bf 1008}, 031 (2010)
  [arXiv:1005.4439 [hep-th]].

\bibitem{Seiberg:1999xz} 
  N.~Seiberg and E.~Witten,
  JHEP {\bf 9904}, 017 (1999)
  [hep-th/9903224].

\bibitem{Buchel:2004rr} 
  A.~Buchel,
  Phys.\ Rev.\ D {\bf 70}, 066004 (2004)
  [hep-th/0402174].

\bibitem{Kleban:2004bv} 
  M.~Kleban, M.~Porrati and R.~Rabadan,
  JHEP {\bf 0508}, 016 (2005)
  [hep-th/0409242].

\bibitem{Maldacena:1998uz} 
  J.~M.~Maldacena, J.~Michelson and A.~Strominger,
  JHEP {\bf 9902}, 011 (1999)
  [hep-th/9812073].


\bibitem{birmingham} D. Birmingham and S. Mokhtari, \textit{Stability of Topological Black Holes}, hep-th/07092388
\bibitem{ishibashi} A. Ishibashi and H. Kodama, \textit{A master equation for gravitational perturbations of maximally symmetric black holes in higher dimensions}, Prog. Theor. Phys. \textbf{110} (2003); hep-th/0305147
\bibitem{Dias} O. Dias, R. Monteiro, H. Reall and J. Santos, \textit{A scalar field condensation instability of rotating anti-de Sitter black holes}; hep-th/10073745
\bibitem{GibbonsHartnoll} Gary Gibbons and Sean A. Hartnoll, \textit{Gravitational instability in higher dimensions}, hep-th/0206202


\bibitem{Wang:2001tk} 
  B.~Wang, E.~Abdalla and R.~B.~Mann,
  Phys.\ Rev.\ D {\bf 65}, 084006 (2002)
  [hep-th/0107243].

\bibitem{Koutsoumbas:2008yq} 
  G.~Koutsoumbas, E.~Papantonopoulos and G.~Siopsis,
  Class.\ Quant.\ Grav.\  {\bf 26}, 105004 (2009)
  [arXiv:0806.1452 [hep-th]].

\bibitem{Aros:2002te} 
  R.~Aros, C.~Martinez, R.~Troncoso and J.~Zanelli,
  Phys.\ Rev.\ D {\bf 67}, 044014 (2003)
  [hep-th/0211024].
  
\bibitem{vasiliev} M. A. Vasiliev, \textit{Higher spin gauge theories: Star product and AdS space}; hep-th/9910096
\bibitem{Sezgin:2002rt} 
  E.~Sezgin and P.~Sundell,
  Nucl.\ Phys.\ B {\bf 644}, 303 (2002)
  [Erratum-ibid.\ B {\bf 660}, 403 (2003)]
  [hep-th/0205131].


\bibitem{klebanovpolyakov} I.R. Klebanov, A.M. Polyakov, \textit{AdS Dual of the Critical O(N) Vector Model}; hep-th/0210114

\bibitem{Giombi:2012ms} 
  S.~Giombi and X.~Yin,
  J.\ Phys.\ A {\bf 46}, 214003 (2013)
  [arXiv:1208.4036 [hep-th]].



\bibitem{BF1} P. Breitenlohner and D. Z. Freedman, \textit{Stability In Gauged Extended Supergravity} Annals Phys. \textbf{144} (1982) 249;
\bibitem{BF2} P. Breitenlohner and D. Z. Freedman, \textit{Positive Energy In Anti-De Sitter Backgrounds And Gauged Extended Supergravity} Phys. Lett. B \textbf{115} (1982) 197.
\bibitem{lightstates} S. Banerjee, S. Hellerman, J. Maltz, S. H. Shenker, \textit{Light States in Chern-Simons Theory Coupled to Fundamental Matter}; hep-th/1207.4195
\bibitem{LargeNCSMatter} O. Aharony, S. Giombi, G. Gur-Ari, J. Maldacena, R. Yacoby, \textit{The Thermal Free Energy in Large N Chern-Simons-Matter Theories}; hep-th/1211.4843
\bibitem{dSTopology} S. Banerjee, A. Belin, S. Hellerman, A. Lepage-Jutier, A. Maloney, D. Radicevic, S. Shenker, \textit{Topology of Future Infinity in dS/CFT}; hep-th/1306.6629

\bibitem{Gopakumar:1998ki} 
  R.~Gopakumar and C.~Vafa,
  Adv.\ Theor.\ Math.\ Phys.\  {\bf 3}, 1415 (1999)
  [hep-th/9811131].
  
\bibitem{Maldacena:2012sf} 
  J.~Maldacena and A.~Zhiboedov,
  Class.\ Quant.\ Grav.\  {\bf 30}, 104003 (2013)
  [arXiv:1204.3882 [hep-th]].

\bibitem{Aharony:2012nh} 
  O.~Aharony, G.~Gur-Ari and R.~Yacoby,
  JHEP {\bf 1212}, 028 (2012)
  [arXiv:1207.4593 [hep-th]].


\bibitem{GurAri:2012is} 
  G.~Gur-Ari and R.~Yacoby,
  JHEP {\bf 1302}, 150 (2013)
  [arXiv:1211.1866 [hep-th]].


\bibitem{Aharony:2012ns} 
  O.~Aharony, S.~Giombi, G.~Gur-Ari, J.~Maldacena and R.~Yacoby,
  JHEP {\bf 1303}, 121 (2013)
  [arXiv:1211.4843 [hep-th]].

\bibitem{Jain:2013py} 
  S.~Jain, S.~Minwalla, T.~Sharma, T.~Takimi, S.~R.~Wadia and S.~Yokoyama,
  JHEP {\bf 1309}, 009 (2013)
  [arXiv:1301.6169 [hep-th]].



\bibitem{ads5_s5} H.J. Kim, L.J. Romans and P. van Nieuwenhuizen, \textit{Mass spectrum of chiral ten-dimensional $N=2$ supergravity on $S^5$} Physical Review D \textbf{32} (1985) 389
\bibitem{cp3} B.E.W Nilsson and C.N. Pope, \textit{Hopf fibration of eleven-dimensional supergravity}, Class. Quant. Grav. \textbf{I} (1984) 499-515
\bibitem{mq} O. DeWolfe, D. Freedman, S. Gubser, G. Horowitz and I. Mitra, \textit{Stability of $AdS_p \times M_q$ compactifications without supersymmetry}, Physical Review D \textbf{65} (2002) 064033
\bibitem{klebanov} I.R. Klebanov and E. Witten, \textit{Superconformal field theory on three-branes at a Calabi-Yau singularity} Nucl. Phys. \textbf{B536} (1998) 199
\bibitem{t11} A. Ceresole, G. Dall'Agata, R. D'Auria and S. Ferrara, \textit{Spectrum of Type IIB Supergravity on $AdS_5\times T^{11}$: Predictions on $\mathcal{N}=1$ SCFT's} hep-th/9905226
\bibitem{gubser} S. Gubser, \textit{Einstein Manifolds and Conformal Field Theories}, hep-th/9807164
\bibitem{Baumann:2010sx} 
  D.~Baumann, A.~Dymarsky, S.~Kachru, I.~R.~Klebanov and L.~McAllister,
  JHEP {\bf 1006}, 072 (2010)
  [arXiv:1001.5028 [hep-th]].



















\end{thebibliography}
\end{document}